\documentclass[10pt,superscriptaddress,nofootinbib,notitlepage,tightenlines,twocolumn, pra]{revtex4-2}

\usepackage{makecell}
\usepackage{amsthm,amsmath,amssymb}
\usepackage{mathrsfs}
\usepackage{amsthm}
\usepackage{amsmath}
\usepackage{amssymb}
\usepackage{graphicx}
\usepackage{epstopdf}
\usepackage{url}
\usepackage{float}
\usepackage{bm}
\usepackage{ifthen}
\usepackage[usenames,dvipsnames]{color}
\usepackage{mathrsfs}
\usepackage[colorlinks=true,citecolor=blue,urlcolor=black]{hyperref}
\usepackage{subfigure}
\usepackage{multirow}
\usepackage{makecell}

\newcommand{\be}{\begin{equation}}
\newcommand{\ee}{\end{equation}}

\newcommand\xrowht[2][0]{\addstackgap[.5\dimexpr#2\relax]{\vphantom{#1}}}
\newcommand{\sket}[1]{{\ensuremath{\lvert#1\rangle}}}
\newcommand{\lket}[1]{{\ensuremath{\left\lvert#1\right\rangle}}}
\newcommand{\ket}[1]{\if@display\lket{#1}\else\sket{#1}\fi}

\usepackage{stackengine}
\newcommand{\sbra}[1]{{\ensuremath{\langle#1\rvert}}}
\newcommand{\lbra}[1]{{\ensuremath{\left\langle#1\right\rvert}}}
\newcommand{\bra}[1]{\if@display\lbra{#1}\else\sbra{#1}\fi}

\newcommand{\sbraket}[2]{{\ensuremath{\langle#1\rvert#2\rangle}}}
\newcommand{\lbraket}[2]{{\ensuremath{\left\langle#1\!\left\rvert\vphantom{#1}#2\right.\!\right\rangle}}}
\newcommand{\braket}[2]{\if@display\lbraket{#1}{#2}\else\sbraket{#1}{#2}\fi}

\newcommand{\sketbra}[2]{{\ensuremath{\lvert #1\rangle\!\langle #2\rvert}}}
\newcommand{\lketbra}[2]{{\ensuremath{\left\lvert #1\right\rangle\!\!\left\langle #2\right\rvert}}}
\newcommand{\ketbra}[2]{\if@display\lketbra{#1}{#2}\else\sketbra{#1}{#2}\fi}

\theoremstyle{plain}

\theoremstyle{definition}

\begin{document}

\title{Advantages of Asynchronous Measurement-Device-Independent Quantum Key Distribution in Intercity Networks}

\author{Yuan-Mei Xie}
\author{Jun-Lin Bai}
\author{Yu-Shuo Lu}
\author{Chen-Xun Weng}
\author{Hua-Lei Yin}\email{hlyin@nju.edu.cn}
\author{Zeng-Bing Chen}\email{zbchen@nju.edu.cn}
\affiliation{National Laboratory of Solid State Microstructures and School of Physics, Collaborative Innovation Center of Advanced Microstructures, Nanjing University, Nanjing 210093, China}


\begin{abstract}
The new variant of measurement-device-independent quantum key distribution (MDI-QKD),  called asynchronous MDI-QKD or mode-pairing MDI-QKD, offers similar
repeater-like rate-loss scaling but has the advantage of simple technology implementation  by exploiting an innovative post-measurement pairing technique.
We herein present  an evaluation of the practical aspects of decoy-state asynchronous MDI-QKD. To determine its effectiveness, we analyze the optimal  method  of decoy-state calculation and examine the impact of asymmetrical channels and multi-user networks. Our simulations show that, under realistic conditions, aynchronous MDI-QKD can furnish the highest key rate with MDI security as compared to other QKD protocols over distances ranging from 50 km to 480 km. At fiber distances of 50 km and 100 km, the key rates attain 6.02 Mbps and 2.29 Mbps respectively, which are sufficient to facilitate real-time one-time-pad video encryption. Our findings indicate that experimental implementation of asynchronous MDI-QKD in intercity networks can be both practical and efficient.
\end{abstract}

\maketitle
\section{Introduction}
Quantum key distribution (QKD)~\cite{bennett2014quantum, ekert1991quantum}  enables remote two parties to share secret keys protected from eavesdropping by the laws of physics. In the past forty years, QKD has
achieved rapid development in terms of secret key rates~\cite{scarani2009security, xu2020secure,pirandola2020advances,grunenfelder2023fast}, transmission distance~\cite{gobby2004quantum,
frohlich2017long,boaron2018secure} and network deployment~\cite{Peev_2009,sasaki2011field,dynes2019cambridge,chen2021integrated}. 
Although the security of QKD has been proven in theory, the imperfections of realistic devices lead to various security loopholes~\cite{zhao2008quantum,lydersen2010hacking,tang2013source}, especially in detection~\cite{lydersen2010hacking}.
 
Fortunately, measurement-device-independent (MDI) QKD is proposed~\cite{lo2012measurement}, which assumes an untrusted intermediate node to perform two-photon Bell state measurements, thus solving all security issues at the detection side~\cite{braunstein2012side}. Extensive work demonstrates the potential of MDI-QKD, including  experimental breakthroughs~\cite{Liu2013Experimental,rubenok2013real,experiment2014mdi,yin2016measurement,comandar2016quantum,yin2017experimental,woodward2021gigahertz}, on-chip implementations~\cite{semenenko2020chip,zheng2021heterogeneously,wei2020high}, and continuous theoretical developments~\cite{curty2014finite,yin2014long,zhou2016making,wang17measurement,wang2019asymmetric,azuma2022quantum}. Moreover, users in a MDI-QKD network can share expensive detectors, and the topology of MDI-QKD is naturally suitable for   deployment in star-type networks. Additionally, side-channel-secure QKD  has recently been experimentally realized, which is not only MDI but also immune to potential source imperfections~\cite{zhang2022experimental,gu2022experimental}.
However, the key rates of most forms of QKD  are fundamentally bounded by the secret key capacity of repeaterless QKD~\cite{pirandola2009direct,takeoka2014fundamental,pirandola2017fundamental,das2021universal} due to photon loss in the channel. A rigorous theorem,
the absolute repeaterless secret key capacity
(SKC$_0$), expresses this limit as $R=-\log_2(1-\eta)$ ~\cite{pirandola2017fundamental}, i.e., the key rate $R$ scales linearly with the channel transmittance $\eta$. Despite some progress in overcoming this bound~\cite{jiang2009quantum,munro2012quantum,azuma2015allrepeaters,azuma2015all}, such devices remain elusive.

Twin-field (TF) QKD~\cite{Lucamarini2018overcoming} and its variants~\cite{ma2018phase,wang2018twin,yin2019measurement,lin2018simple,cui2019twin,curty2019simple} are proposed to break this bound. The protocols make the untrusted intermediate node use Bell state measurements based on single-photon interference, rather than two-photon interference. Numerous works have advanced  theory with finite key analysis~\cite{maeda2019repeaterless, yin2019finite, jiang2019unconditional,curras2021tight}. Ref.~\cite{Li2021long} applies entangled coherent state sources as untrusted relays  to  further increase the transmission distance of TF-QKD by reducing the signal-to-noise ratio at the measurement nodes. Several
experimental achievements have shown the performance
of twin-field QKD over large loss~\cite{minder2019experimental,zhong2019proof,wang2019beating,liu2019experimental,fang2020implementation,chen2020sending,liu2021field,zhong2021proof,chen2021twin,clivati2022coherent,Pittaluga2021600-km,wang2022twin,li2022twin,zhou2023quantum},
and the maximum distance of TF-QKD has
been experimentally increased to 830 kilometers~\cite{wang2022twin}.
The idea of single-photon interference  has also   been implemented in  device independent QKD~\cite{xie2021overcoming}.
Nonetheless, as TF-QKD  requires stable long-distance single-photon interference, phase-tracking and phase-locking techniques are indispensable~\cite{Lucamarini2018overcoming}. These techniques are complicated and expensive, and usually impose a negative impact on the system performance. For example, phase tracking technology requires  sending strong reference light, which reduces the effective clock frequency of the quantum signal and increases background noise~\cite{fang2020implementation,chen2020sending,Pittaluga2021600-km,wang2022twin}.

Recently, the new variant~\cite{xie2022breaking, zeng2022mode} of MDI-QKD, called asynchronous MDI-QKD~\cite{xie2022breaking} (also called mode-pairing MDI-QKD~\cite{zeng2022mode}), is proposed. It asynchronously pairs two successful clicks within a long pairing time to establish two-photon Bell state, thereby breaking SKC$_0$. Asynchronous MDI-QKD is highly practical and  has a noteworthy advantage over TF-QKD in intercity-distance quantum communications, owing to its implementation simplicity and performance.  Several exciting experiments have successfully verified the superior performance of asynchronous MDI-QKD with accessible technology. Ref.~\cite{zhu2023experimental} realizes the experiment with a maximal distance of 407 km  without global phase locking. Ref.~\cite{zhou2022experimental} demonstrates the first asynchronous MDI-QKD  that overcomes SKC$_0$ without global phase tracking and extends the maximal distance to 508 km. 
However, before asynchronous MDI-QKD can be applied in real life, many issues of practicality necessitate resolution, such as identifying the optimal number of decoy states, determining the optimal calculation method of decoy states, and assessing the performance in asymmetric channels and networks.

In this work, we  address these issues by introducing  the joint-constraints technique~\cite{jiang2021higher} and new methods for phase error rate estimation  to enable higher-rate asynchronous MDI-QKD. 
By employing the three-intensity protocol alongside an additional \textit{click filtering}  operation---which is the known best choice for performance---we simulate the key rate of asynchronous MDI-QKD in multi-user networks.   For a network of five users, asynchronous MDI-QKD result in the key rates of all links surpassing the secret key capacity. Furthermore, using a 4 GHz repetition rate  system~\cite{wang2022twin}, secret key
rates of 6.02 Mbps, 2.29 Mbps, and 0.31 Mbps can be achieved at fiber distances of 50 km, 100 km, and 200 km, respectively. Asynchronous MDI-QKD can achieve the highest key rate in the range of 170 to 480 km, compared with decoy-state QKD~\cite{hwang2003quantum,wang2005beating,lo2005decoy} and TF-QKD~\cite{Lucamarini2018overcoming}. More importantly, our work provides conceptual differences between asynchronous MDI-QKD and its synchronous version (original time-bin  MDI-QKD)~\cite{ma2012alternative} in Sec.~\ref{sec_conclusion}. Asynchronous MDI-QKD holds the most promising potential as a solution for intercity-distance quantum communication in the future, owing to its minimal detector requirements and absence of strong light feedback.

\section{Protocol description} 
Here, we consider an asymmetric asynchronous MDI-QKD protocol using three-intensity setting, which is similar to the protocol described in Ref.~\cite{zhou2022experimental}, but offers the option to use \textit{click  filtering} or not. The intensity
of each laser pulse is randomly set to one of the three intensities
$\mu_{a(b)}$ (signal), $\nu_{a(b)}$ (decoy) and $o_{a(b)}$ (vacuum), and the intensities satisfy $\mu_{a(b)}>\nu_{a(b)}>o_{a(b)}=0$. 
A successful click is obtained when one and only one detector clicks in a time bin, and we refer to $(k_{a}|k_{b})$ as a successful click when Alice sends intensity $k_a$ and Bob sends $k_b$. The 
notation $[k_{a}^{\rm tot}, k_{b}^{\rm tot}]$ indicates an asynchronous coincidence where the combined intensity in the two time-bins Alice (Bob) sent is $k_{a}^{\rm tot}$ ($k_{b}^{\rm tot}$). The details of the protocol are presented as follows.
 
\textit{1. Preparation.} For each time bin, Alice chooses a phase value   $\theta_a=2\pi M_{a}/M$ with $M_{a}\in \{0,1,...,M-1\}$
at random. Then, she selects an intensity choice $k_a\in \{\mu_a,\nu_a,o_a\}$   with probabilities $p_{\mu_a}$, $ p_{\nu_a}$, and $ p_{o_a}=1-p_{\mu_a}-p_{\nu_a}$, respectively. Alice prepares a weak laser pulse $\ket{e^{\textbf{i}\theta_a}\sqrt{k_a}}$ based on the chosen values. Similarly, Bob
prepares a weak coherent pulse $\ket{e^{\textbf{i}\theta_b}\sqrt{k_b}}$~($k_b\in \{\mu_b,\nu_b,o_b\}$). Finally, Alice and Bob send their optical pulses to Charlie via  the quantum channel. 

\textit{2. Measurement.}  For each time bin, Charlie performs a first-order interference measurement on the two received pulses, and he publicly announces whether a successful click is obtained and which detector  ($D_{L}$ or $D_{R}$) clicked. The first two steps will be repeated $N$ times.

\textit{3. Coincidence pairing.} 
The clicks that Alice and Bob retained for further processing depend on whether \textit{click  filtering} is applied. If they  perform  \textit{click filtering},
Alice (Bob) announces whether she (he) applied the decoy intensity $\nu_{a}$ ($\nu_{b}$) to the pulse sent for each event. Then they discard clicks $(\mu_{a}|\nu_{b})$ and $(\nu_{a}|\mu_{b})$, and keep all other clicks. Otherwise, they keep all clicks. 

For all kept clicks, Alice and Bob always pair a click with the nearest one within a time interval $T_c$ to form a successful coincidence. They discard the lone click that failed to find a partner within $T_c$.  For each coincidence, Alice (Bob) computes the total intensity used between the two time bins $k_a^{\rm tot}$ ($k_b^{\rm tot}$) and  the phase differences between the early ($e$) and late ($l$) time bins, $\varphi_{a(b)}=\theta_{a(b)}^{l}-\theta_{a(b)}^{e}$.

\textit{4. Sifting.} Alice and Bob announce  their computational results and then discard the data if $k_a^{\rm tot}=\mu_a+\nu_a$ or $k_b^{\rm tot}= \mu_b+\nu_b$. When there is a \textit{click filtering} operation, we define $\tilde{k}_{a(b)}=\mu_{a(b)}$; otherwise, we define $\tilde{k}_{a(b)} \in \{\mu_{a(b)},\nu_{a(b)}\}$.
For $[\tilde{k}_a, \tilde{k}_b]$ coincidence,  Alice 
 (Bob) extracts a $\boldsymbol{Z}$-basis bit 0 (1)  if she (he) sends $\tilde{k}_{a(b)}$ in the early  time bin and $o_{a(b)}$ in the late time bin. Otherwise,  Alice (Bob) extracts an opposite bit.
Note that we use four intensity
groups ($[\mu_a, \mu_b], [\mu_a, \nu_b], [\nu_a, \nu_b], [\nu_a, \mu_b]$) for the key generation when \textit{click filtering} is not applied, while existing MDI-QKD protocols typically use only one intensity group.
For $[2\nu_a, 2\nu_b]$ and $[2\mu_a, 2\mu_b]$ coincidences, Alice and Bob calculate the relative phase difference $\varphi_{ab}=(\varphi_{a}-\varphi_{b})\mod 2 \pi$. They extract an $\boldsymbol{X}$-basis bit 0 if $\varphi_{ab} = 0$ or $\pi$.  Afterwards, Bob flips his bit value, if $\varphi_{ab} = 0$ and both detectors clicked, or $\varphi_{ab} = \pi$ and the same detector clicked twice.  The coincidence with other phase differences is discarded. 

\textit{5. Parameter estimation.} Alice and Bob group their data into different sets $\mathcal{S}_{[k_{a}^{\rm tot},k_{b}^{\rm tot}]}$ and count the corresponding number $n_{[k_{a}^{\rm tot},k_{b}^{\rm tot}]}$. By using all the raw data they have obtained, Alice and Bob estimate the necessary parameters to calculate the key rate. They estimate the  number of vacuum events, $s_0^z$, the number of single-photon pair events in  the $\boldsymbol{Z}$ basis, $s_{11}^z$, the bit error rate of the single-photon pairs in the $\boldsymbol{X}$ basis, $e_{11}^x$, and the phase error rate associated with the single-photon pair events in  the $\boldsymbol{Z}$ basis, $\phi^{z}_{11}$.

\textit{6. Key distillation.} Alice and Bob perform an error correction step that reveals at most $\lambda_{\rm EC}$ bits of information.  Under the condition of passing the checks in the error correction  
and privacy amplification steps, a $\varepsilon_{\rm tot}$-secure key of length~\cite{zhou2022experimental,portmann2022security}
\begin{equation}
\begin{aligned}
\ell=&  \underline{s}_{0}^z+\underline{s}_{11}^z\left[1-H_2\left(\overline{\phi}_{11}^z\right)\right]-\lambda_{\rm EC} 
\\ 
& \log_2\frac{2}{\varepsilon_{\rm cor}}-2\log_2\frac{2}{\varepsilon'\hat{\varepsilon}}-2\log_2\frac{1}{2\varepsilon_{\rm PA}} ,\label{eq_keyrate}
\end{aligned}
\end{equation}
can be extracted, where $\underline{x}$ and $\overline{x}$ are the lower and
upper bounds of the observed value $x$, respectively; $H_2(x)=-x \log_2 x-(1-x) \log_2(1-x)$ is the binary Shannon entropy function.  Using the entropic uncertainty relation~\cite{zhou2022experimental}, the total secure coefficient  $\varepsilon_{\rm tot}=2(\varepsilon'+2\varepsilon_e+\hat{\varepsilon})+\varepsilon_0+\varepsilon_1+\varepsilon_\beta+\varepsilon_{\rm PA}+\varepsilon_{\rm cor}$, where $\varepsilon_{\rm cor}$ is the failure probability of error correction; $\varepsilon_{\rm PA}$ is the failure probability of privacy amplification; $\hat{\varepsilon}$ and $\varepsilon'$ are the coefficients while using a chain-rule for smooth entropies; $\varepsilon_0$,  $\varepsilon_1$ and $\varepsilon_\beta$ are the failure probabilities for estimating the terms of $s_0^z$, $s_{11}^z$, and $e_{11}^x$, respectively.

\section{The key rate formula} 
In the following description,
let $x^*$ be the expected value of $x$. In the asynchronous MDI-QKD protocol, $[\tilde{k}_a, \tilde{k}_b]$ coincidence can be used to generate keys. Since the binary
Shannon entropy function is concave, we can correct errors for each group $[\tilde{k}_a, \tilde{k}_b]$ separately to reduce the consumption of information, which does not affect the security of the protocol. Hence the amount of information consumed in error correction can be written as
\begin{equation}
\begin{aligned}
\lambda_{\rm{EC}}&=\sum\limits_{\tilde{k}_a,\tilde{k}_b}\left[ n_{[\tilde{k}_a,\tilde{k}_b]}fH_2\left( E_{[\tilde{k}_a,\tilde{k}_b]}\right)\right],
\end{aligned}\label{eq:lambda}
\end{equation}
where $f$ is the error correction efficiency and $E_{[\tilde{k}_a,\tilde{k}_b]}$ is the bit error rate of $[\tilde{k}_a,\tilde{k}_b]$ coincidence. Because vacuum states contain no information about their bit values, in the asymmetric case we can separately extract higher-valued vacuum components in each group $[\tilde{k}_a, \tilde{k}_b]$  to obtain higher key rates. 
The total number of vacuum components in the $\boldsymbol{Z}$  basis can be given by
\begin{equation}
\begin{aligned}
\underline{s}_{0}^{z*}= \sum\limits_{\tilde{k}_a,\tilde{k}_b}   \max\left\{
\frac{e^{-\tilde{k}_a } p_{[\tilde{k}_a,\tilde{k}_b]}}{p_{[o_a,\tilde{k}_b]}}\underline{n}_{[o_a,\tilde{k}_b]}^{ *}, \frac{e^{-\tilde{k}_b}p_{[\tilde{k}_a,\tilde{k}_b]}}{p_{[\tilde{k}_a,o_b]}}\underline{n}_{[\tilde{k}_a,o_b]}^{ *}\right\}.
\end{aligned}\label{s0z_start}
\end{equation}
Here $p_{[k_a^{\rm tot},k_b^{\rm tot}]}$ is the probability that $[k_a^{\rm tot},k_b^{\rm tot}]$ coincidence  occurs given the coincidence event, which is 
\begin{equation}
\begin{aligned}
 p_{[k_a^{\rm tot},k_b^{\rm tot}]}=  \sum\limits_{k_a^e+k_a^l= k_a^{\rm tot}}\sum\limits_{k_b^e+k_b^l=k_b^{\rm tot}}  
  \frac{p_{k_a^e}p_{k_b^e}}{p_s} \frac{p_{k_a^l}p_{k_b^l}}{p_s}, \label{Eq:ptot}
\end{aligned}
\end{equation}
apart from $p_{[2\nu_a,2\nu_b]}$ because of the phase matching condition in the $\boldsymbol{X}$ basis, which is
\begin{equation}
	\begin{aligned}
	p_{[2\nu_a,2\nu_b]}=\frac{2}{M}\frac{p_{\nu_a}p_{\nu_b}}{p_{s}} \frac{p_{\nu_a}p_{\nu_b}}{p_{s}}.
	\end{aligned} 
\end{equation}
When  \textit{click filtering} is not applied, $p_s=1$, otherwise $p_s=1-p_{\mu_a}p_{\nu_b}-p_{\nu_a}p_{\mu_b}$. 

Next, we need to estimate the number and phase error rate of the single-photon pairs in the 
 $\boldsymbol{Z}$ basis, $s_{11}^z$ and $\phi_{11}^z$. Because the density matrices of single-photon pairs are identical in the $\boldsymbol{Z}$ and $\boldsymbol{X}$ bases,  the expected ratio of different intensity settings is the same for all single-photon pairs~\cite{lo2012measurement,zhou2016making}; namely,
\begin{equation}
\begin{aligned}
\frac{\underline{s}_{11}^{z*}}{\underline{s}_{11}^{x*}}=\frac{t_{11}^{z*}}{t_{11}^{x*}}= \frac{\sum_{\tilde{k}_a,\tilde{k}_b}\left(\tilde{k}_a\tilde{k}_be^{-\tilde{k}_a-\tilde{k}_b} p_{[\tilde{k}_a,\tilde{k}_b]}\right)}{4\nu_a\nu_be^{-2\nu_a-2\nu_b}p_{[2\nu_a,2\nu_b]}},\label{s11_relation}\\
\end{aligned}
\end{equation}
where $t_{11}^z$ represents the number of errors of the  single-photon pairs in the $\boldsymbol{Z}$, while $s_{11}^x$ and $t_{11}^x$ denote the number of single-photon pairs and their corresponding error count  in  $[2\nu_a,2\nu_b]$ coincidence, respectively. 

Then we estimate the lower bound of ${s}_{11}^{z*}$ using the decoy-state method~\cite{hwang2003quantum,wang2005beating,lo2005decoy}, which can be given by 
\begin{equation}	
\begin{aligned}
 \underline{s}_{11}^{z*}=&  \frac{\sum_{\tilde{k}_a,\tilde{k}_b}\left(\tilde{k}_a\tilde{k}_be^{-\tilde{k}_a-\tilde{k}_b} p_{[\tilde{k}_a,\tilde{k}_b]}\right)}{\nu_a\nu_b\mu_a\mu_b(\mu'-\nu')}\\&\times \left[\mu_a\mu_b\mu' \left(e^{\nu_a+\nu_b }\frac{\underline{n}_{[\nu_a ,\nu_b] }^{*}}{p_{[\nu_a,\nu_b]}}-e^{\nu_b}\frac{\overline{n}_{[o_a,\nu_b] }^*}{p_{[o_a,\nu_b] }}\right.\right.\\
 &\left.\left.- 
	e^{\nu_a }\frac{\overline{n}_{[\nu_a ,o_b] }^{*}}{p_{[\nu_a,o_b]}}
	+\frac{\underline{n}_{[o_a,o_b] }^{*}}{p_{[o_a,o_b]}} \right)
\right.\\ 
	&	 -\nu_a\nu_b\nu' \left(e^{\mu_a+\mu_b}\frac{\overline{n}_{[\mu_a,\mu_b] }^{*}}{p_{[\mu_a,\mu_b] }}-e^{\mu_b }\frac{\underline{n}_{[o_a,\mu_b] }^*}{p_{[o_a,\mu_b] }}\right.\\
 &\left.\left.- 
	e^{\mu_a}\frac{\underline{n}_{[\mu_a, o_b] }^{*}}{p_{[\mu_a,o_b]}}
	+ \frac{\underline{n}_{[o_a,o_b] }^{*}}{p_{[o_a,o_b] }} \right)	\right], 
\end{aligned}\label{eq_decoy_Y11}
\end{equation}
where
\begin{equation}
   \begin{cases}
   \mu'=\mu_a,\quad \nu'=\nu_a,& \mbox{if} \quad \frac{\mu_a}{\mu_b}
   \le \frac{\nu_a}{\nu_b}, \\
   \mu'=\mu_b, \quad\nu'=\nu_b,&\mbox{if}  \quad \frac{\mu_a}{\mu_b}
   > \frac{\nu_a}{\nu_b}.
   \end{cases}
  \end{equation}
We can use the technique of joint constraints~\cite{jiang2021higher} to obtain the tighter estimated value of ${s}_{11}^{z*}$. The details of the analytic results of joint constraints are shown in Appendix~\ref{joint_constraint_append}.  
Then we can obtain the lower bound of ${s}_{11}^{x*}$ with Eq.~\eqref{s11_relation}.

\begin{table}[b!]
\centering
\caption{Simulation parameters. Here $\eta_d=\eta_d^L=\eta_d^R$, $p_d=p_d^L=p_d^R$, and $\eta_d^L~(\eta_d^R)$ and $p_d^L~(p_d^R)$ are the detection efficiency and the dark count rate of the detector $D_L~(D_R)$, respectively; $\alpha$ denotes the attenuation coefficient of the fiber; $\omega_{\rm fib}$ is the fiber phase drift rate; $E_{\rm HOM}$ is the interference misalignment error rate; $f$  is the error correction efficiency; $\Delta \nu$ is the laser frequency difference;   and $\epsilon$ is the failure probability considered in the error verification and finite data analysis.}\label{tab1}
\begin{tabular}[b]{@{\extracolsep{0pt}}cccccccc}
	\hline
	\hline
	$\eta_{d} $  & $p_{d} $ & $\alpha$ & $\omega_{\rm fib}$ & $E_{\rm HOM}$ &  $f$  &$\Delta\nu$ & $\epsilon$\\
	\hline\xrowht{7pt}
	$80\%$ & 0.1 Hz   & $0.16$ dB/km  & $5900$ rad/s   & 0.04 &	 $1.1$  & 10 Hz & $10^{-10}$\\
	\hline
	\hline
\end{tabular}
\end{table}

The upper bound of the single-photon pair errors in the $\boldsymbol{X}$ basis is 
\begin{align}
\overline{t}_{11}^{x}= m_{[2\nu_a,2\nu_b]}-\underline{m}_{[2\nu_a,2\nu_b]}^{0},\label{phi_11z_start}
\end{align}
where $m_{[2\nu_a,2\nu_b]}$ is the observed error bit number in the $\boldsymbol{X}$ basis, and $m_{[2\nu_a,2\nu_b]}^{0}$ is the  error bit number in the $\boldsymbol{X}$ basis given that at least one of Alice and Bob sends vacuum component. The lower bound of the expected value $m_{[2\nu_a,2\nu_b]}^{0*}$ can be given by
\begin{equation}
\begin{aligned}\label{mo}
\underline{m}_{[2\nu_a,2\nu_b]}^{0*}=&\frac{e^{-2\nu_a}p_{[2\nu_a,2\nu_b]}}{2p_{[o_a,2\nu_b]}}\underline{n}_{[o_a,2\nu_b]}^{*}+\frac{e^{-2\nu_b}p_{[2\nu_a,2\nu_b]}}{2p_{[2\nu_a,o_b]}}\underline{n}_{[2\nu_a,o_b]}^{*}\\
&- \frac{e^{-2\nu_a-2\nu_b}p_{[2\nu_a,2\nu_b]}}{2p_{[o_a,o_b]}}\overline{n}_{[o_a,o_b]}^{*}.
\end{aligned}
\end{equation}
Using similar arguments, we obtain the tighter value of $\underline{m}_{[2\nu_a,2\nu_b]}^{0*}$ under the joint constraints~\cite{jiang2021higher}.

For single-photon pairs, the expected value of the
phase error rate in the $\boldsymbol{Z}$ basis equals the expected value
of the bit error rate in the $\boldsymbol{X}$ basis, and the error rate $\overline{e}_{11}^x=\overline{t}_{11}^x/{\underline{s}_{11}^{x}}$. There are two methods for estimating $\overline{\phi}_{11}^{z}$. The first method involves using the random sampling method to estimate $\overline{\phi}_{11}^{z}$ from  $\overline{e}_{11}^{x}$~\cite{zhou2022experimental}. Explicitly~\cite{yin2020tight},
\begin{equation}
\begin{aligned}
\overline{\phi}_{11}^{z}=& \overline{e}_{11}^x	+\gamma\left(\underline{s}_{11}^z,\underline{s}_{11}^x,\overline{e}_{11}^x,\varepsilon_e\right),\\ 
\end{aligned}\label{Randomswr}
\end{equation}
where
\begin{equation}
\gamma^{U}(n,k,\lambda,\epsilon)=\frac{\frac{(1-2\lambda)AG}{n+k}+
	\sqrt{\frac{A^2G^2}{(n+k)^2}+4\lambda(1-\lambda)G}}{2+2\frac{A^2G}{(n+k)^2}},
\end{equation}
with $A=\max\{n,k\}$ and $G=\frac{n+k}{nk}\ln{\frac{n+k}{2\pi nk\lambda(1-\lambda)\epsilon^{2}}}$. 

On the other hand, following Ref.~\cite{zhou2016making}, an alternative approach involves using the observed values of  $t_{11}^z$ to estimate the upper bound for $\phi_{11}^z$. Specifically,
\begin{equation}
	\begin{aligned}
		\overline{\phi}_{11}^{z}=& \frac{\overline{t}_{11}^{z}}{\underline{s}_{11}^z},\label{phi_11z_stop}
	\end{aligned} 
\end{equation}
where the upper bound of $t_{11}^{z}$ and the lower bound of $s_{11}^{z}$ can be estimated by $\overline{t}_{11}^{z*}$ and $\underline{s}_{11}^{z*}$  with the Chernoff bound (see Eqs.~\eqref{chernoff1}~and~\eqref{chernoff2} in Appendix~\ref{statistical}). We can calculate $\overline{t}_{11}^{z*}$ with Eq.~\eqref{s11_relation} and 
\begin{align}
\overline{t}_{11}^{x*}= m_{[2\nu_a,2\nu_b]}^*-\underline{m}_{[2\nu_a,2\nu_b]}^{0*}.
\end{align}

\begin{figure}[t!]
\centering
\includegraphics[width=8.6cm]{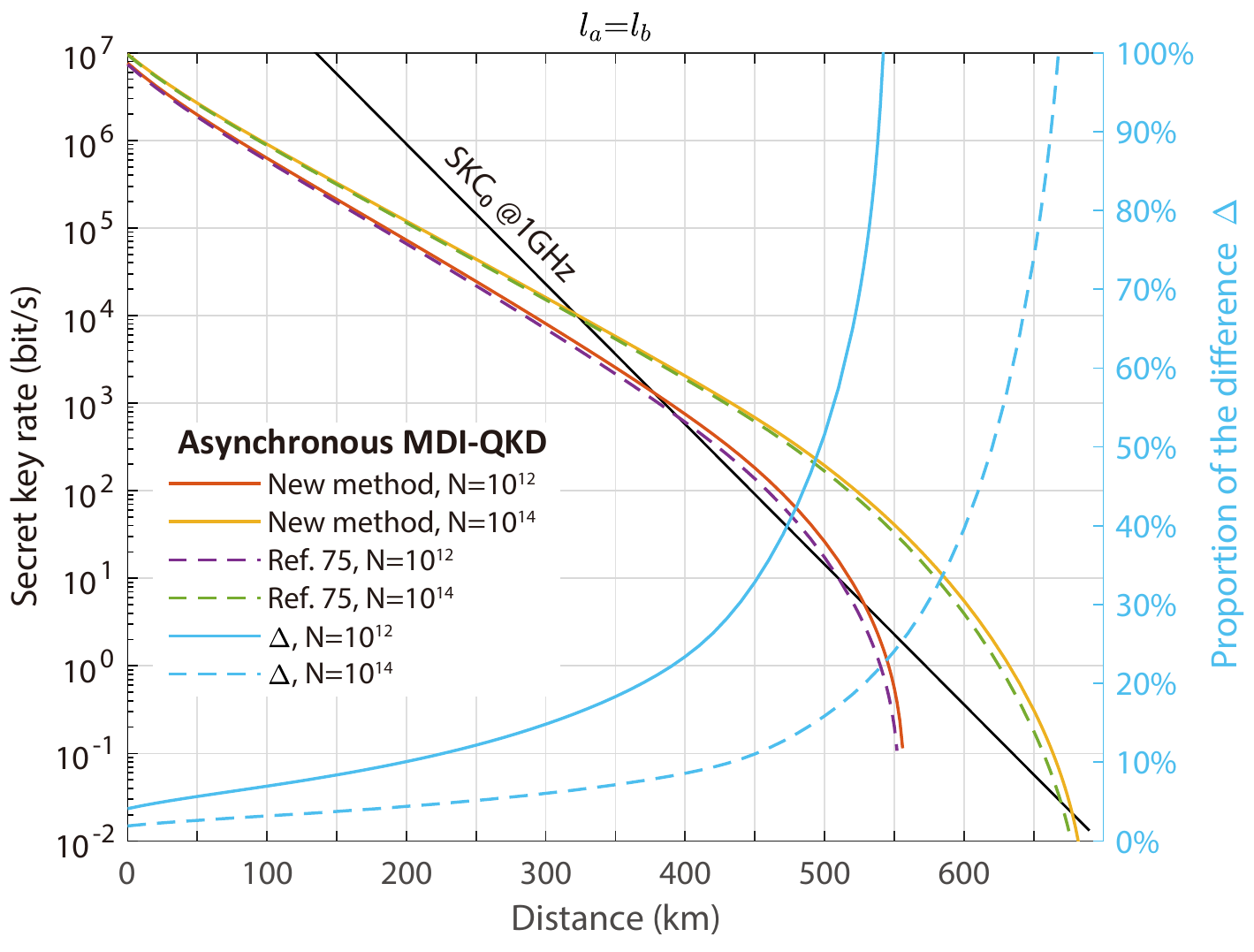}
\caption{Secret key rates of the three-intensity  asynchronous MDIQKD protocol with \textit{click filtering} using different phase error rate estimation methods. Here $l_a~(l_b)$ is the distance between  Alice (Bob) and Charlie. The horizontal axis represents the total transmission distance  $l = l_a + l_b$. The  relative difference between the secret key rates of the new  method  $R_{\rm new}$ and that of the original method  $R_{\rm ori}$ is shown with the $y$ axis on the right. $\Delta=(R_{\rm new}-R_{\rm ori})/R_{\rm ori}.$} The numerical results here show that the new phase error rate estimation method has a notable advantage.\label{fig:amdi_sy_f_newvsold} 
\end{figure}

 \begin{figure*}
\centering
\includegraphics [width=17.6cm]{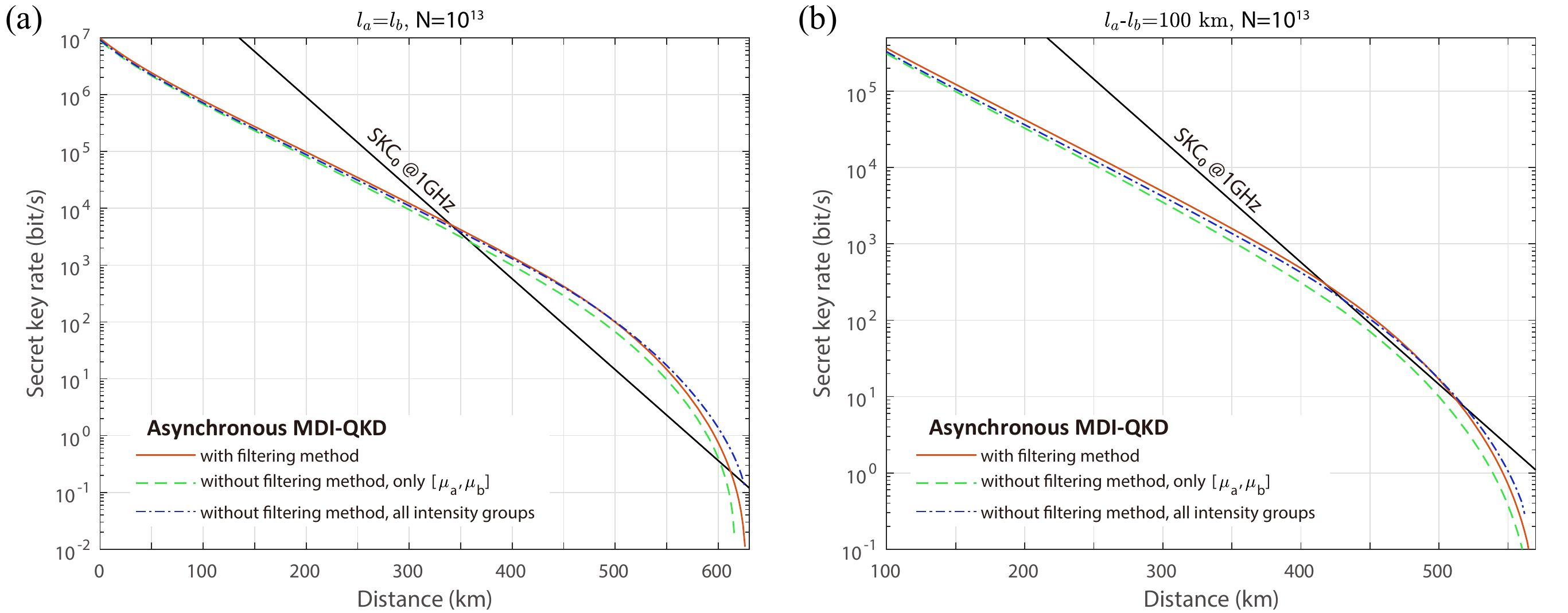}
\caption{ Comparison of the secret key rates of  asynchronous MDIQKD with  and without
\textit{click filtering} under two types of
channels: (a) symmetric channel $l_a=l_b$ and  (b)  asymmetric channels $l_a-l_b=100$ km. The horizontal axis represents the total transmission distance $l = l_a + l_b$. }\label{fig:amd_nfvsf_10hz}
\end{figure*}

 \section{Performance }

\subsection{Optimal decoy-state method}
For the evaluation, we numerically optimize the secret key
rate $R := \ell F/N$  of
 asynchronous-MDIQKD with Eq.~\eqref{Randomswr} (original method~\cite{zhou2022experimental}) and Eq.~\eqref{phi_11z_stop} (new method), which is shown in
Fig.~\ref{fig:amdi_sy_f_newvsold}. Here $F$ is the system clock frequency. In this work, we set failure parameters $\varepsilon_{\rm{cor}}$,
$\varepsilon'$, $\varepsilon_e$, $\hat{\varepsilon}$, $\varepsilon_{\beta}$, and $\varepsilon_{\rm PA}$ to be the same
value: $\epsilon$.   The experimental parameters are set to the values  used in the state-of-the-art system, as shown in Table~\ref{tab1}.
We denote the distance between Alice (Bob) and Charlie  
by $l_a~(l_b)$.
In Fig.~\ref{fig:amdi_sy_f_newvsold}, we set $F=1$ GHz and $l_a = l_b$, and the source parameters of Alice and Bob are all
the same. The genetic algorithm is exploited to globally search for the optimal value of light
intensities and their corresponding probabilities. The black line is the results of SKC$_0$.  We denote the relative difference between the key rate  of the new method $R_{\rm new}$ and that of the original method $R_{\rm ori}$ as $\Delta=(R_{\rm new}-R_{\rm ori})/R_{\rm ori}$. The results show that as the distance increases, the influence of statistical fluctuations becomes increasingly significant,
and the key rate advantage of the new phase error rate estimation method is also increasing.
For example, at a fiber length of 600 km with $N=10^{14}$, the secret key rate obtained by the new phase
error rate estimation method is approximately  1.49 times that of the  original method.
In the following key rate calculations, we  use the new phase error rate estimation method by default.

 \subsection{Optimal protocol}
 Figure~\ref{fig:amd_nfvsf_10hz} shows a comparison of the secret key rates of asynchronous MDI-QKD with and
without \textit{click filtering} under symmetrical $l_a=l_b$ and asymmetrical channels $l_a-l_b=100$ km. The parameters are
listed in Table~\ref{tab1}.  $F=1$ GHz and $N=10^{13}$ are used. The green dotted line is   results of using only $[\mu_a,\mu_b]$ coincidence to form the secret key without \textit{click filtering}. In the  symmetric channel, Fig.~\ref{fig:amd_nfvsf_10hz}(a), we can see that the  key rate of asynchronous MDI-QKD   with \textit{click filtering} is always  higher than that of asynchronous MDI-QKD without \textit{click filtering} based on $[\mu_a,\mu_b]$ group. This is expected since the filtering operation corresponds to a higher number of valid pairs and smaller statistical fluctuations in the estimation process. And the key rate of  asynchronous MDI-QKD  with \textit{click filtering} is  higher than that of asynchronous MDI-QKD  without \textit{click filtering} based on four intensity groups at short and medium distances. At a fiber length of 300 km, the
secret key rate obtained with  \textit{click filtering} is approximately  1.11 times
the one without \textit{click filtering} based on four intensity groups, and 1.29 times the one based on $[\mu_a,\mu_b]$ group. At longer distances, the effectiveness of click filtering is diminished by a decrease in coincidence pairing efficiency due to less frequent photon clicks. Therefore, in scenarios where click filtering is not utilized, incorporating additional intensity groups ($[\mu_a,\nu_b], [\nu_a,\mu_b], [\nu_a,\nu_b]$) for key generation can lead to higher key rates at longer distances than using click filtering alone.
  The same trend is observed for the  asymmetric channel (Fig.~\ref{fig:amd_nfvsf_10hz}(b)).

\begin{figure}[t]
\centering
\includegraphics[width=8.6cm]{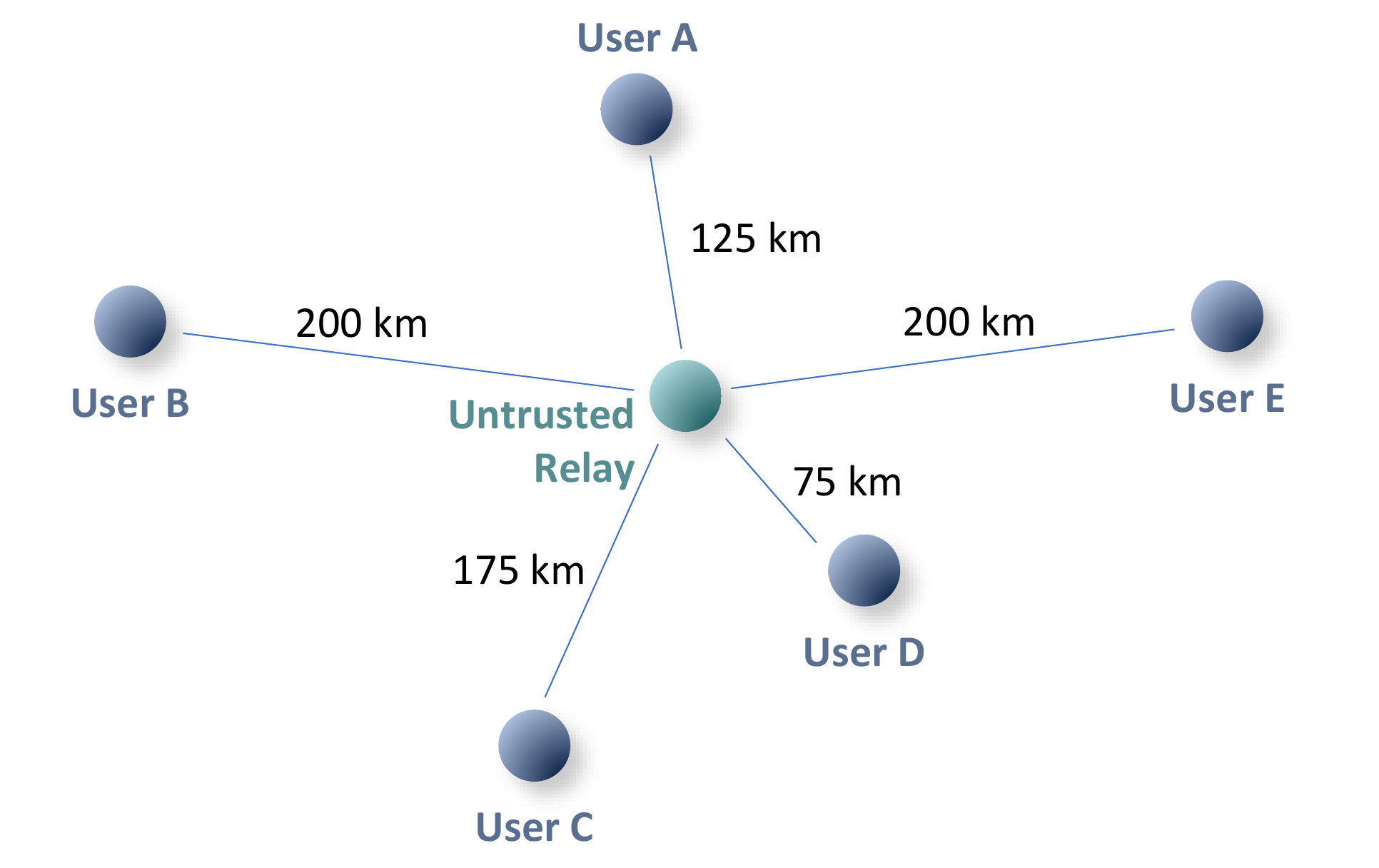}
\caption{Example of a scalable QKD network setup consisting of numerous users who may freely join or leave the network. Each user node has an asymmetric
channel connected to an untrusted relay, through which it can establish a QKD link to others.}\label{amdi_network} 
\end{figure} 

\begin{table*}[t]
\centering
\caption{Simulated secret key rates per second for asynchronous  MDI-QKD,  SNS-QKD with the AOPP method, and PM-QKD in the QKD network shown in Fig.~\ref{amdi_network} using the parameters in Table~\ref{app:tab1}.  The system clock frequency is 4 GHz and the transmission time  is 22 hours. Here, link A-B represents that user A communicates with user B. The sending intensities and corresponding probabilities are selected by the users to obtain the optimal key rate for each link. Note that here we consider a 50\% duty cycle for the TF-type protocols~\cite{minder2019experimental,Pittaluga2021600-km,zhou2023quantum}.}\label{tab3}
	\begin{tabular}[b]{@{\extracolsep{10pt}}c cc ccccc }
        \hline
	\hline	\xrowht{7pt}
 Link  &A-B (A-E) & B-C (C-E)& B-D (D-E)& B-E& A-C & A-D& C-D\\
	\hline\xrowht{7pt}
SKC$_0$ & 5.77  $\times 10^{3}$ & 4.80  $\times 10^{3}$& 1.45  $\times 10^{4}$& 2.30  $\times 10^{3}$ &  1.21 $\times 10^{4}$ & 3.64  $\times 10^{4}$ & 3.03  $\times 10^{4}$ \\
Asynchronous MDI-QKD      &   1.47 $\times 10^{4}$ &   1.36 $\times 10^{4}$& 2.05 $\times 10^{4}$&
9.46 $\times 10^{3}$ &  2.36 $\times 10^{4}$   & 4.04 $\times 10^{4}$&   3.56 $\times 10^{4}$\\
 SNS-QKD (AOPP)  &  1.18  $\times 10^{4}$ &   1.09 $\times 10^{4}$& 1.64 $\times 10^{4}$&
7.53  $\times 10^{3}$ & 1.78 $\times 10^{4}$   & 3.05 $\times 10^{4}$& 2.72 $\times 10^{4}$\\
 PM-QKD &  2.56 $\times 10^{3}$ &   2.40 $\times 10^{3}$ & 3.22 $\times 10^{3}$&
1.71 $\times 10^{3}$ & 4.19 $\times 10^{3}$   & 6.91 $\times 10^{3}$ &  6.01 $\times 10^{3}$\\
	\hline
	\hline
\end{tabular}
\end{table*}

\subsection{Asynchronous MDI-QKD Networks}
We provide a figure about a scalable QKD network setup consisting of numerous users who may freely
join or leave the network in Fig.~\ref{amdi_network}. Each user node has an asymmetric channel connected to an untrusted relay, through which it can establish a QKD link to others. The users will adjust the sending intensities and corresponding probability values so that each link can obtain the optimal key rate.
The experimental parameters used here are listed in Table~\ref{app:tab1}. 

Table~\ref{tab3} shows simulated secret key rates per second for asynchronous MDI-QKD, sending-or-not-sending QKD (SNS-QKD) with actively odd-parity pairing (AOPP)~\cite{jiang2020zigzag},  and phase-matching QKD (PM-QKD)~\cite{zeng2020symmetry} in the QKD intercity network. Assuming a clock rate of 4 GHz~\cite{wang2022twin}  and a transmission time of 22 hours, which corresponds to approximately $3.2\times 10^{14}$ quantum pulses for asynchronous MDI-QKD. We further assume that the quantum transmission duty ratio for the SNS-QKD and PM-QKD systems is 50\%~\cite{minder2019experimental,Pittaluga2021600-km,zhou2023quantum}. Note that duty cycle ratios are lower in many important TF-QKD experiments, for example, the duty ratio at 402 km is 22.4\% in Ref.~\cite{fang2020implementation}, 45\%   in Ref.~\cite{chen2020sending}, and 40\%  in Ref.~\cite{wang2022twin}. The duty cycle has two effects on the key rate. Firstly, the total number of quantum pulses transmitted per second depends on the system clock frequency and the duty cycle. Secondly, the key rate per second is obtained by multiplying the key rate per pulse with the total number of quantum pulses transmitted per second.  We can see that asynchronous MDI-QKD enables the key rates of all links to exceed SKC$_0$. Additionally, asynchronous MDI-QKD  always enjoys  higher secret key rates per clock than SNS-QKD~(AOPP) and PM-QKD.  

\subsection{Practical advantages  of asynchronous MDI-QKD}
\begin{figure}[t]
\centering
\includegraphics[width=8.6cm]{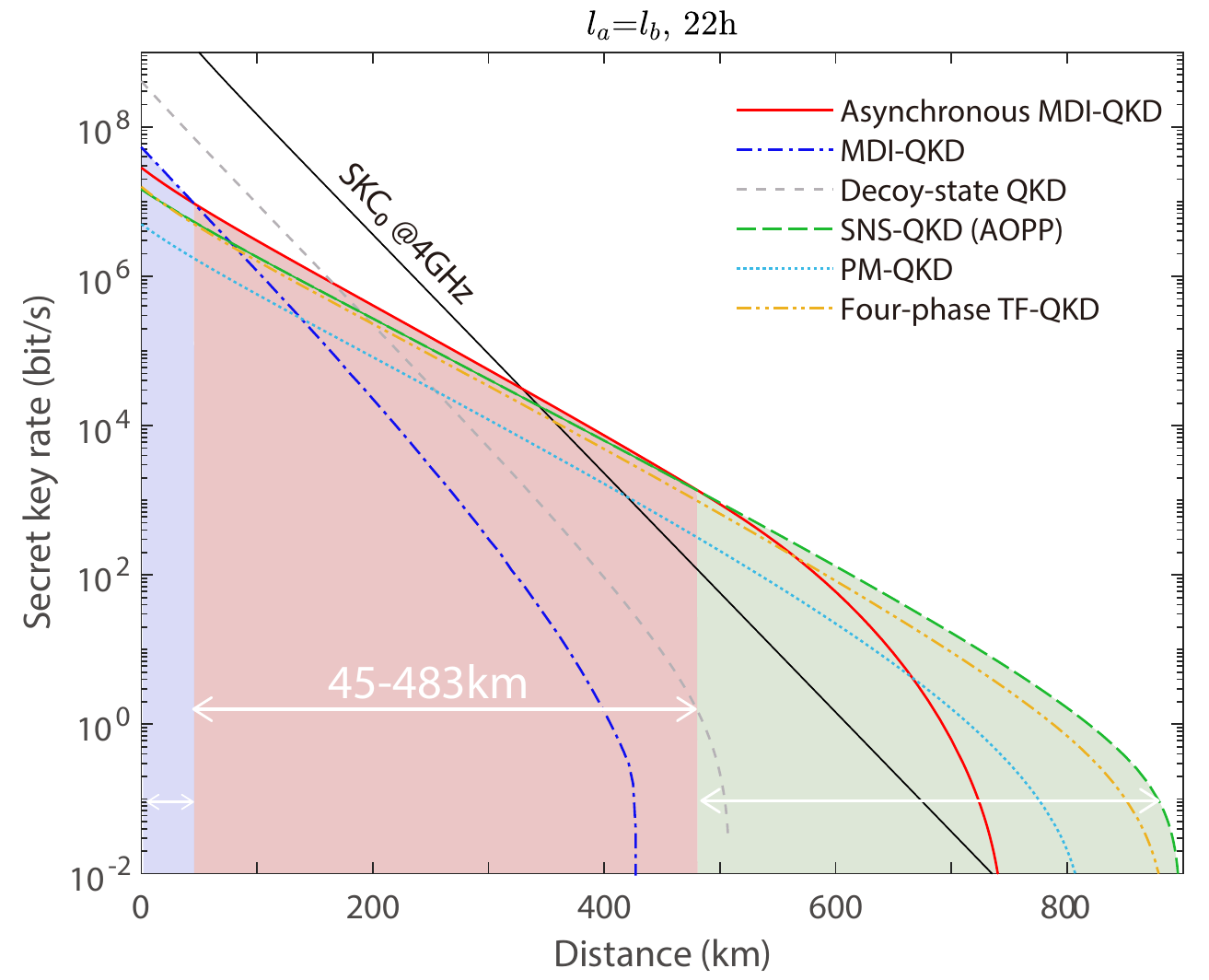}
\caption{Simulated secret key rates for asynchronous  MDI-QKD, original  time-bin MDI-QKD, decoy-state QKD, SNS-QKD with the AOPP method, PM-QKD, and four-phase TF-QKD under the state-of-the-art system. The horizontal axis represents the total transmission distance  $l = l_a + l_b$.}\label{amdi_performance} 
\end{figure}

We simulate the performance of our protocol assuming
a 4 GHz clock rate and 22 hours transmission time.
Figure \ref{amdi_performance} presents the key rate per second versus fiber distance for asynchronous MDI-QKD, together with four-intensity time bin MDI-QKD~\cite{jiang2021higher}, SNS-QKD (AOPP)~\cite{jiang2020zigzag}, PM-QKD~\cite{zeng2020symmetry}, four-phase TF-QKD~\cite{wang2022twin}, and four-intensity decoy-state QKD. 
For SNS-QKD (AOPP), PM-QKD, and four-phase TF-QKD, we set the duty cycle to 50\%,  Charlie's transmission loss at Alice's (Bob's) side to 2 dB, 
the angles of misalignment to $20^{\circ}$, which contributes to an interference error rate of approximately 3\%.
 We assume an insert loss on Bob’s side of 2 dB and a misalignment error rate of $e_m=0.02$ for decoy-state QKD.  The interference misalignment error rate of  decoy-state MDI-QKD is set to 0.04, which corresponds to 27\% error rate in the $\boldsymbol{X}$ basis. Device parameters are shown in Table ~\ref{app:tab1}. The simulation formulas of MDI-QKD and decoy-state QKD are detailed in Appendix~\ref{simulation:MDIQKD} and~\ref{simulation:QKD}, respectively. We also include  SKC$_0$ to prove the repeater-like behavior for asynchronous  MDI-QKD. Simulation shows that the key rate of our protocol surpasses that of the decoy-state QKD protocol when $l>170$ km, and it exceeds SKC$_0$ when $l>330$ km. In the  170-483 km range, the performance of our protocol is better than that of the other five protocols, especially in the range of 200-300 km.  We observe that, in the simulations, the key rates of decoy-state QKD surpass those of original time bin MDI-QKD due to the influence of the dark rate and finite key analysis. At short distances (less than 45km), asynchronous MDI-QKD has a  slightly lower key rate compared to the original time-bin MDI-QKD. This is attributed to the stronger light intensity of the signal state in the original MDI-QKD, approaching 1, which results in a higher number of single-photon pairs in the $\boldsymbol{Z}$ basis. In
Table~\ref{tab4}, we present the bits-per-second (bps) values of asynchronous MDI-QKD at various typical distances, employing device parameters identical to those employed in Fig.~\ref{amdi_performance}. Our protocol can generate secret keys  rate of 0.31 Mbps at a fiber length of 200 km, thereby rendering it adequate for secure key-demanding applications such as real-time one-time-pad secure audio encryption in intra- and inter-urban areas.

\begin{table*}[!htp]
\centering
\caption{ The secret key rates of the three-intensity  asynchronous MDI-QKD protocol with \textit{click filtering}. Here the fiber loss is 0.16 dB/km; the clock rate is 4 GHz; the dark count rates is 0.1 Hz; and the detection efficiency is  $\eta_d=80\%$.}\label{tab4}
\begin{tabular}[b]{@{\extracolsep{9pt}}c cccccc }
        \hline
	\hline
 	\xrowht{8pt}
 	Data size &     $ 10^{12}$ & 5 $\times 10^{12}$ &  $10^{13}$ &  $10^{13}$ &   5 $\times 10^{13}$ &      5 $\times 10^{13}$   \\
	\cline{2-7}  \noalign{\smallskip}  Distance (km) & 
      50  & 100  & 150     & 200  & 250  & 300   \\
	\hline\xrowht{8pt}
Secret key rate & 6.02	 Mbps	& 2.29 Mbps& 	855.40  kbps & 305.05 kbps  &  129.60 kbps &	46.671 kbps \\
	\hline
	\hline
\end{tabular}
\end{table*}

\section{Discussion and conclusion}
\label{sec_conclusion}	

Here, we  point out two conceptual differences between asynchronous MDI-QKD and original  MDI-QKD.

\romannumeral 1.  In original MDI-QKD, the total number of sent pluses allows for a direct measurement of the ``gain'', while the ``yield'' of single-photon pairs in the Z and X bases can be estimated using decoy-state methods~\cite{curty2014finite}. However, in asynchronous MDI-QKD, where post-measurement coincidence pairing is utilized, there is no concept of the total sent pair number. Therefore, the terms ``gain'' and ``yield'' are not applicable.

\romannumeral 2. In asynchronous MDI-QKD, the terms "three-intensity" and "four-intensity" refer to the number of light intensities used, and the intensities at different bases after pairing are associated. Specifically, in three-intensity asynchronous MDI-QKD, there are two intensities in each of the $\boldsymbol{Z}$ and $\boldsymbol{X}$ bases after coincidence pairing. These intensities are associated as follows: in the $\boldsymbol{Z}$ basis, the intensities are $\mu$ and $\nu$, while in the $\boldsymbol{X}$ basis, the intensities are $2\mu$ and $2\nu$, and the non-basis intensity is 0. In contrast, in the original three-intensity MDI-QKD, there is only one intensity in the $\boldsymbol{Z}$ basis.

In the original MDI-QKD protocol, an important idea is to consider the double-scanning method~\cite{jiang2021higher}. We have applied the double-scanning method to asynchronous MDI-QKD. 
The derivation details of double-scanning are shown in Appendix~\ref{double_scanning_append}. However, numerical results show that the method does not work for the three-intensity  asynchronous MDI-QKD protocol~\cite{xiecode}. We remark that this phenomenon may be caused by the above two important characteristics. In asynchronous MDI-QKD, the number of single-photon pairs in the $\boldsymbol{Z}$  basis can be accurately estimated using $\boldsymbol{Z}$-basis data, without the need for inefficient $\boldsymbol{X}$-basis data.  Additionally, there is a correlation between the intensities used to estimate the number of the $\boldsymbol{Z}$-basis single-photon pairs and the intensities used to estimate the $\boldsymbol{X}$-basis phase error rate in asynchronous MDI-QKD. In contrast, the intensity and decoy-state estimation in the $\boldsymbol{Z}$ and $\boldsymbol{X}$ bases are independent in original MDI-QKD, which makes double scanning an effective strategy.

Furthermore, in the original MDI-QKD protocol, we can improve the performance of the protocol by increasing the number of decoy states, such as four-intensity MDI-QKD~\cite{zhou2016making}. We also have calculated the key rate of the four-intensity asynchronous MDI-QKD protocol, in which the intensity of each laser pulse is randomly set to one of the four intensities
$\mu_{a(b)}$ (signal), $\omega_{a(b)}$ (decoy 1), $\nu_{a(b)}$ (decoy 2) and $o_{a(b)}$ (vacuum), and the intensities satisfy $\mu_{a(b)}>\omega_{a(b)}>\nu_{a(b)}>o_{a(b)}=0$. The detailed calculation of the protocol is presented in Appendix~\ref{four_AMDI}. Comparing  secret key rates of the three-intensity and  four-intensity asynchronous MDI-QKD protocol with \textit{click filtering}, we find that the optimal key rates for the four-intensity decoy-state method are nearly equal to the results for the three-intensity decoy-state method~\cite{xiecode}.  We remark that this situation is also due to the correlation between intensities at different bases. Therefore, the three-intensity asynchronous MDI-QKD protocol is a good trade-off between the performance of key rates and  the ease of implementation. 

In this work, we have presented an analysis of the practical aspects of  asynchronous MDI-QKD. We have provided refined decoy-state methods that enable higher-rate asynchronous MDI-QKD. The numerical results of different asynchronous MDI-QKD protocols demonstrate that the three-intensity protocol, with a  \textit{click filtering} operation, can provide a favorable balance between performance and ease of implementation.  We have introduced the decoy-state method for the asymmetric situation, which permits the direct application of our protocol to asynchronous MDI-QKD experiments with asymmetric channels. Our work also provides  important  insights into asynchronous MDI-QKD: the decoy-state analysis for the $\boldsymbol{Z}$ and $\boldsymbol{X}$ bases of asynchronous MDI-QKD are correlated, rendering the introduction of double scanning and additional decoy states ineffective for key rate improvement. With its superior performance and straightforward design, asynchronous MDI-QKD holds strong potential in future quantum networks spanning 200 to 400 km.  We anticipate the application of the asynchronous concept to MDI multiparty quantum communication tasks, such as
quantum conference key agreement~\cite{fu2015long}, quantum
secret sharing~\cite{fu2015long}, and quantum digital signatures~\cite{yin2022experimental}.

\par
\begin{center}
\textbf{ACKNOWLEDGMENTS} 
\end{center}
The authors acknowledge Z. Yuan and L. Zhou for the insightful discussions.
 This work has been supported by the National Natural Science Foundation of China (No. 12274223), the Natural Science Foundation of Jiangsu Province (No. BK20211145), the Fundamental Research Funds for the Central Universities (No. 020414380182), the Key Research and Development Program of Nanjing Jiangbei New Area (No. ZDYD20210101),  the Program for Innovative Talents and Entrepreneurs in Jiangsu (No. JSSCRC2021484), and the Program of Song Shan Laboratory (Included in the management of Major  Science and Technology Program of Henan Province) (No. 221100210800-02).

\appendix

\section{Analytic results of joint constraints}\label{joint_constraint_append}
Here, we introduce the joint-constraints method to bound tighter values. Without loss of generality, we take Eq.~\eqref{eq_decoy_Y11} as an example. Similar operations can be applied to other parameters. We can rewrite Eq.~\eqref{eq_decoy_Y11} as
\begin{align}
\underline{s}_{11}^{z*}\geq	& \frac{\sum_{\tilde{k}_a,\tilde{k}_b}\left(\tilde{k}_a\tilde{k}_be^{-\tilde{k}_a-\tilde{k}_b} p_{[\tilde{k}_a,\tilde{k}_b]}\right)}{\nu_a\nu_b\mu_a\mu_b(\mu'-\nu')} \left(\underline{S}_1^*-\overline{S}^*_2\right),
\end{align}
where
\begin{equation}
\begin{aligned}
S_1=&\mu_a\mu_b\mu'e^{\nu_a+\nu_b }\frac{n_{[\nu_a ,\nu_b] } }{p_{[\nu_a,\nu_b]}}+\nu_a\nu_b\nu'e^{\mu_b }\frac{n_{[o_a,\mu_b] }}{p_{[o_a,\mu_b] }}\\
&+\nu_a\nu_b\nu'e^{\mu_a}\frac{n_{[\mu_a, o_b] } }{p_{[\mu_a,o_b]}}+(\mu_a\mu_b\mu'-\nu_a\nu_b\nu')\frac{n_{[o_a,o_b] } }{p_{[o_a,o_b]}},
  \end{aligned}
  \end{equation}
and
\begin{equation}
\begin{aligned}
S_2=&\nu_a\nu_b\nu' e^{\mu_a+\mu_b}\frac{n_{[\mu_a,\mu_b] } }{p_{[\mu_a,\mu_b] }}+\mu_a\mu_b\mu'e^{\nu_b}\frac{n_{[o_a,\nu_b] }^*}{p_{[o_a,\nu_b] }}\\
&+ \mu_a\mu_b\mu' e^{\nu_a }\frac{n_{[\nu_a ,o_b] } }{p_{[\nu_a,o_b]}}.
  \end{aligned}
  \end{equation}
For $\underline{S}_1^*$, we define
\begin{equation}
\begin{aligned}
S_1:=& a_1  \gamma_1 +a_2  \gamma_2+a_3  \gamma_3+a_4  \gamma_4,
  \end{aligned}
  \end{equation}
where $a_1=\frac{\mu_a\mu_b\mu'e^{\nu_a+\nu_b }}{p_{[\nu_a,\nu_b]}}$, $\gamma_1=n_{[\nu_a ,\nu_b] }$, $a_2=\frac{\nu_a\nu_b\nu'e^{\mu_b }}{p_{[o_a,\mu_b] }}$, $\gamma_2=n_{[o_a,\mu_b]}$, $a_3=\frac{\nu_a\nu_b\nu'e^{\mu_a}}{p_{[\mu_a,o_b]}}$, $\gamma_3=n_{[\mu_a, o_b] }$, $a_4=\frac{\mu_a\mu_b\mu'-\nu_a\nu_b\nu' }{p_{[o_a,o_b]}}$, $\gamma_4=n_{[o_a,o_b]}$.
Denoting  $\{b_1,b_2,b_3,b_4\}$
as the ascending order of $\{a_1,a_2,a_3,a_4\}$, and  $\xi_1$, $\xi_2$, $\xi_3$, $\xi_4$ as the 
corresponding rearrange of $\{\gamma_1,\gamma_2,\gamma_3,\gamma_4\}$  according to the ascending
order of  $\{a_1,a_2,a_3,a_4\}$, then we have the lower bound of $S^*_1$~\cite{jiang2021higher}:
\begin{equation}
\begin{aligned}
\underline{S}^*_1:=& b_1  \underline{(\xi_1+\xi_2+\xi_3+\xi_4)}^* +(b_2-b_1)  \underline{(\xi_2+\xi_3+\xi_4)}^*\\
&+(b_3-b_2)  \underline{(\xi_3+\xi_4)}^*+(b_4-b_3)  \underline{\xi}^*_4.
  \end{aligned}
  \end{equation}
For $\overline{S}_2^*$, we define
\begin{equation}
\begin{aligned}
S_2:=& c_1  \kappa_1 +c_2  \kappa_2+c_3  \kappa_3+c_4  \kappa_4
  \end{aligned}
  \end{equation}
where $a_1=\frac{\mu_a\mu_b\mu'e^{\nu_a+\nu_b }}{p_{[\nu_a,\nu_b]}}$, $\gamma_1=n_{[\nu_a ,\nu_b] }$, $a_2=\frac{\nu_a\nu_b\nu'e^{\mu_b }}{p_{[o_a,\mu_b] }}$, $\gamma_2=n_{[o_a,\mu_b]}$, $a_3=\frac{\nu_a\nu_b\nu'e^{\mu_a}}{p_{[\mu_a,o_b]}}$, $\gamma_3=n_{[\mu_a, o_b] }$, $a_4=\frac{\mu_a\mu_b\mu'-\nu_a\nu_b\nu' }{p_{[o_a,o_b]}}$, $\gamma_4=n_{[o_a,o_b]}$.
Denoting  $\{d_1,d_2,d_3\}$
as the ascending order of $\{c_1,c_2,c_3\}$, and  $\chi_2$, $\chi_3$, as the 
corresponding rearrange of $\{\kappa_1,\kappa_2,\kappa_3\}$ according to the ascending
order of  $\{c_1,c_2,c_3\}$, then we have the upper bound of $S^*_2$~\cite{jiang2021higher}:
\begin{equation}
\begin{aligned}
\overline{S}^*_2:=& d_1\times \overline{(\chi_1+\chi_2+\chi_3)}^* +(d_2-d_1)\times \overline{(\chi_2+\chi_3)}^*\\
&+(d_3-d_2)\times \overline{\chi}^*_3.
  \end{aligned}
  \end{equation}

\section{Decoy-state estimation with the double-scanning method}\label{double_scanning_append}
Here we apply the double-scanning method to asynchronous MDI-QKD.
Using the decoy-state method, we can estimate
the lower bound of the number of single-photon pairs in the $\boldsymbol{X}$ basis 
\begin{equation}
\begin{aligned}\label{eq:s11x}
\underline{s}_{11}^{x*}=\frac{e^{-2\nu_a-2\nu_b}p_{[2\nu_a,2\nu_b]}}{\mu_a\mu_b(\tilde{\mu}'-\tilde{\nu}')} (\underline{S}^{+*}-\overline{S}^{-*}-\overline{H}^*), 
\end{aligned}
\end{equation}
where
\begin{equation}
   \begin{cases}
   \tilde{\mu}'=2\mu_a,\quad \tilde{\nu}'=2\nu_a, & \mbox{if} \quad \frac{\mu_a}{\mu_b}
   \le \frac{\nu_a}{\nu_b}, \\
   \tilde{\mu}'=2\mu_b, \quad \tilde{\nu}'=2\nu_b,& \mbox{if}  \quad \frac{\mu_a}{\mu_b}
   > \frac{\nu_a}{\nu_b},
   \end{cases}
  \end{equation}
and
\begin{equation}
        \begin{aligned}      S^{+*}=&\mu_a\mu_b\tilde{\mu}' e^{2\nu_a+2\nu_b }\frac{n_{[2\nu_a, 2\nu_b] }^{*}}{p_{[2\nu_a,2\nu_b]}}+\nu_a\nu_b\tilde{\nu}' e^{2\mu_b }\frac{\underline{n}_{[o_a,2\mu_b] }^*}{p_{[o_a,2\mu_b] }}\\
        &+ \nu_a\nu_b\tilde{\nu}' 
	e^{2\mu_a}\frac{\underline{n}_{[2\mu_a, o_b] }^{*}}{p_{[2\mu_a,o_b]}},\\
S^{-*}=&\nu_a\nu_b\tilde{\nu}' e^{2\mu_a+2\mu_b}\frac{\overline{n}_{[2\mu_a,2\mu_b] }^{*}}{p_{[2\mu_a,2\mu_b] }}+\nu_a\nu_b\tilde{\nu}' \frac{\underline{n}_{[o_a,o_b] }^{*}}{p_{[o_a,o_b] }},\\
H^*=&\mu_a\mu_b\tilde{\mu}'\left(e^{2\nu_b}\frac{\underline{n}_{[o_a,2\nu_b]}^{*}}{p_{[o_a,2\nu_b]}}+e^{2\nu_a}\frac{\underline{n}_{[2\nu_a,o_b]}^{*}}{p_{[2\nu_a,o_b]}}-\frac{\overline{n}_{[o_a,o_b]}^{*}}{p_{[o_a,o_b]}}\right).\\
        \end{aligned}
        \end{equation} 
The upper bound of the bit error rate of single-photon pairs in the $\boldsymbol{X}$ basis $e_{11}^{x*}$ satisfies
\begin{equation}
\begin{aligned}\label{eq:e11x}
\overline{e}^{x*}_{11}=& \frac{1}{\mu_a\mu_b\tilde{\mu}'e^{2\nu_a+2\nu_b}\underline{s}_{11}^{x*}}\left(\mu_a\mu_b\tilde{\mu}'e^{2\nu_a+2\nu_b}\frac{\underline{m}_{[2\nu_a,2\nu_b]}^{*}}{p_{[2\nu_a,2\nu_b]}}-\frac{H}{2}\right).
\end{aligned}
\end{equation}
Denote $\tilde{n}_{[2\nu_a,2\nu_b]}=n_{[2\nu_a,2\nu_b]}-m_{[2\nu_a,2\nu_b]}$. We can divide the effective 
$[2\nu_a,2\nu_b]$ coincidence into two kinds of events, the right effective events whose total number is $\tilde{n}_{[2\nu_a,2\nu_b]}$, and the wrong effective events whose total number is $m_{[2\nu_a,2\nu_b]}$.
 Denote $M=\mu_a\mu_b\tilde{\mu}'e^{2\nu_a+2\nu_b}\underline{m}_{[2\nu_a,2\nu_b]}^{*}/p_{[2\nu_a,2\nu_b]}$. We can rewrite Eq.~\eqref{eq:s11x} as
 \begin{equation}
\begin{aligned}\label{eq:s11x_new}
\underline{s}_{11}^{x*}=\frac{e^{-2\nu_a-2\nu_b}p_{[2\nu_a,2\nu_b]}}{\mu_a\mu_b(\tilde{\mu}'-\tilde{\nu}')} (\underline{\tilde{S}}^{+*}-\overline{S}^{-*}+\underline{M}^*-\overline{H}^*) ,
\end{aligned}
\end{equation}
where
\begin{equation}
        \begin{aligned}      \tilde{S}^{+*}=&\mu_a\mu_b\tilde{\mu}' e^{2\nu_a+2\nu_b }\frac{\tilde{n}_{[2\nu_a, 2\nu_b] }^{*}}{p_{[2\nu_a,2\nu_b]}}+\nu_a\nu_b\tilde{\nu}' e^{2\mu_b }\frac{n_{[o_a,2\mu_b] }^*}{p_{[o_a,2\mu_b] }}\\
        &+ \nu_a\nu_b\tilde{\nu}' 
	e^{2\mu_a}\frac{n_{[2\mu_a, o_b] }^{*}}{p_{[2\mu_a,o_b]}},\\
S^{-*}=&\nu_a\nu_b\tilde{\nu}' e^{2\mu_a+2\mu_b}\frac{n_{[2\mu_a,2\mu_b] }^{*}}{p_{[2\mu_a,2\mu_b] }}+\nu_a\nu_b\tilde{\nu}' \frac{n_{[o_a,o_b] }^{*}}{p_{[o_a,o_b] }},\\
H^*=&\mu_a\mu_b\tilde{\mu}'\left(e^{2\nu_b}\frac{n_{[o_a,2\nu_b]}^{*}}{p_{[o_a,2\nu_b]}}+e^{2\nu_a}\frac{n_{[2\nu_a,o_b]}^{*}}{p_{[2\nu_a,o_b]}}-\frac{n_{[o_a,o_b]}^{*}}{p_{[o_a,o_b]}}\right).\\
        \end{aligned}
        \end{equation} 
For each group $(H,M)$, we can calculate $e^{x*}_{11}$ with  Eqs.~\eqref{eq:e11x} and \eqref{eq:s11x_new}
\begin{equation}
\begin{aligned}
\overline{e}^{x*}_{11}
=\frac{(\tilde{\mu}'-\tilde{\nu}')(M-H/2)}  {\tilde{\mu}' (S^+-S^-+M-H)}.
\end{aligned}
\end{equation}
By scanning $(H,M)$~\cite{jiang2021higher}, we can get the worst case for  $e^{x*}_{11}$, i.e.,
\begin{alignat}{2}
\max\quad & e_{11}^{x*}&{}& \\
\mbox{s.t.}\quad
&\underline{H} \leq H\leq \overline{H}, \\ \nonumber
&\underline{M} \leq M\leq \overline{M}. \\ \nonumber
\end{alignat}
With the formulas in Eqs.~\eqref{eq:lambda},  \eqref{s0z_start}, ~\eqref{s11_relation}, ~\eqref{eq_decoy_Y11}, and \eqref{phi_11z_stop}, we can get the final key rate.

\section{Four-intensity asynchronous MDI-QKD protocol}\label{four_AMDI}
Here, we provide the decoy-state method for four-intensity asynchronous MDI-QKD with \textit{click filtering}.
The core difference in the parameter estimation steps between  the four-intensity protocol and the three-intensity protocol is to estimate the lower bound of the number of single-photon pairs in the  $\boldsymbol{Z}$ basis. In the four-intensity protocol with \textit{click filtering}, $s_{11}^{z*}$ is bounded by
\begin{equation}	
\begin{aligned}
 \underline{s}_{11}^{z*}=&  \frac{\sum_{\tilde{k}_a,\tilde{k}_b}\left(\tilde{k}_a\tilde{k}_be^{-\tilde{k}_a-\tilde{k}_b} p_{[\tilde{k}_a,\tilde{k}_b]}\right)}{\nu_a\nu_b\omega_a\omega_b(\omega'-\nu')} \\&\times\left[\omega_a\omega_b\omega' \left(e^{\nu_a+\nu_b }\frac{\underline{n}_{[\nu_a ,\nu_b] }^{*}}{p_{[\nu_a,\nu_b]}}-e^{\nu_b}\frac{\overline{n}_{[o_a,\nu_b] }^*}{p_{[o_a,\nu_b] }}\right.\right.\\
 &\left.\left.- 
	e^{\nu_a }\frac{\overline{n}_{[\nu_a ,o_b] }^{*}}{p_{[\nu_a,o_b]}}
	+\frac{\underline{n}_{[o_a,o_b] }^{*}}{p_{[o_a,o_b]}} \right)
\right.\\ 
	&	 -\nu_a\nu_b\nu' \left(e^{\omega_a+\omega_b}\frac{\overline{n}_{[\omega_a,\omega_b] }^{*}}{p_{[\omega_a,\omega_b] }}-e^{\omega_b }\frac{\underline{n}_{[o_a,\omega_b] }^*}{p_{[o_a,\omega_b] }}\right.\\
 &\left.\left.- 
	e^{\omega_a}\frac{\underline{n}_{[\omega_a, o_b] }^{*}}{p_{[\omega_a,o_b]}}
	+ \frac{\underline{n}_{[o_a,o_b] }^{*}}{p_{[o_a,o_b] }} \right)	\right], \label{Y11_decoy_2}
\end{aligned}
\end{equation}
where
\begin{equation}
   \begin{cases}
   \omega'=\omega_a,\quad \nu'=\nu_a& \mbox{if} \quad \frac{\omega_a}{\omega_b}
   \le \frac{\nu_a}{\nu_b}, \\
   \omega'=\omega_b, \quad\nu'=\nu_b&\mbox{if}  \quad \frac{\omega_a}{\omega_b}
   > \frac{\nu_a}{\nu_b},
   \end{cases}
  \end{equation}
and $p_{[k_a^{\rm tot},k_b^{\rm tot}]}$ is defined in Eq.~\eqref{Eq:ptot}. When \textit{click filtering} is not applied, $p_s=1$, otherwise $p_s=1-p_{\mu_a}p_{\omega_b}-p_{\mu_a}p_{\nu_b}-p_{\omega_a}p_{\mu_b}-p_{\omega_a}p_{\nu_b}-p_{\nu_a}p_{\mu_b}-p_{\nu_a}p_{\omega_b}$. Similarly, we use the technique of joint constraints to get the tight estimated value of ${s}_{11}^{z*}$. The calculation of the remaining parameter values can directly utilize Eqs.~\eqref{eq:lambda},  \eqref{s0z_start}, and  \eqref{phi_11z_start}~-~\eqref{phi_11z_stop}.

\begin{table*}[!htp]
\centering
\caption{ List of experimental parameters used in numerical simulations. The spectral filtering loss results from the use of dense-wavelength-division-multiplexing in dual-band TF-type QKD  implementations~\cite{Pittaluga2021600-km,zhou2023quantum}, whereas asynchronous MDI-QKD does not require the dual-band method. Additionally, the asynchronous MDI-QKD system and the decoy-state QKD system do not require a reference pulse, allowing their duty cycle for quantum transmission to be 100\%.} \label{app:tab1}
\begin{tabular}[b]{@{\extracolsep{4pt}}c cccc }
        \hline
	\hline
 	\xrowht{8pt}
 	  &  Asynchronous MDI-QKD & Decoy-state QKD &  SNS-QKD \& PM-QKD &    Four-phase TF-QKD   \\
	\hline\xrowht{8pt}
Fiber loss & 0.16 dB/km 	  &  0.16 dB/km 	& 	 0.16 dB/km 	   &  0.16 dB/km 	\\
Charlie loss & ---&2 dB  & --- &---\\
Detector efficiency & $80\%$  & $80\%$& $80\%$ & $80\%$\\
Dark count rate & 0.1 Hz & 0.1 Hz & 0.1 Hz & 0.1 Hz\\
Spectral filtering loss & 0 dB & 0 dB & \makecell[c]{2 dB at Alice-Charlie \\ 2 dB at Bob-Charlie} & \makecell[c]{2 dB at Alice-Charlie \\ 2 dB at Bob-Charlie}  \\
Duty Cycle & 100 & 100 & 50 & 50\\
Laser frequency difference & 10 Hz &---&---&---\\
Drift rates & $5.9\times 10^{3}$ rad/s &--- &---&---\\
Number of phase slices & 16 & --- & 16 & 4\\
\hline
	\hline
\end{tabular}
\end{table*}

\section{Simulation formulas}\label{simu_pb}
The experimental parameters used for performance comparison of these protocols, asynchronous MDI-QKD, decoy-state QKD, SNS-QKD (AOPP), PM-QKD, and four-phase TF-QKD, are listed in Table ~\ref{app:tab1}.
\subsection{Simulation formulas for asynchronous MDI-QKD}

In asynchronous MDI-QKD,  suppose Alice and Bob send intensities $k_a$ and $k_b$ with phase difference $\theta$, the overall gain is given by [Eq. (C22) in Ref.~\cite{zhou2022experimental}]
\begin{equation}
\begin{aligned}
q_{(k_a|k_b)}=&y_{(k_a|k_b)}^LI_0\left(\eta_d^L\sqrt{\eta_a k_a\eta_bk_b}\right)\\
&+y_{(k_a|k_b)}^R I_0\left(\eta_d
^R\sqrt{\eta_a k_a\eta_bk_b}\right) \\
&-2y_{(k_a|k_b)}^Ly_{(k_a|k_b)}^RI_0\left[(\eta_d
^L-\eta_d
^R)\sqrt{\eta_a k_a\eta_bk_b}\right],
\end{aligned} 
\end{equation} 
where $y_{(k_a|k_b)}^{L(R)}=\left(1-p_d^{L(R)}\right)e^{-\frac{\eta_d
^{L(R)}\left(\eta_a k_a+\eta_b k_b\right)}{2}}$; $\eta_d^L~(\eta_d^R)$ and $p_d^L~(p_d^R)$ are the detection efficiency and the dark count rate of the detector $D_L~(D_R)$, respectively;  $\eta_{a}=10^{-\frac{\alpha l_a}{10}}$ and $\eta_{b}=10^{-\frac{\alpha l_b}{10}}$; $I_0(x)$ refers to the zero-order modified Bessel function of the first kind. 

We define $N_{T_{c}}=FT_{c}$ as the number of time bins within time interval $T_{c}$. The total number of valid successful pairing results is [Eq. (C24) in Ref.~\cite{zhou2022experimental}]
\begin{equation}
n_{\rm tot}= \frac{Nq_{\rm tot}}{1+1/q_{T_{c}}},
\end{equation}
where $q_{\rm tot}$ is the probability of having a click event, and $q_{T_{c}}=1-(1-q_{\rm tot})^{N_{T_{c}}}$ is the probability that at least one click event occurs within the time interval $T_{c}$ after a click time bin. When using the matching method without click filtering, $q_{\rm tot}=\sum_{k_a,k_b}p_{k_a}p_{k_b}q_{(k_a|k_b)}$; when using the matching method with click filtering, $q_{\rm tot}=\sum_{k_a,k_b}p_{k_a}p_{k_b}q_{(k_a|k_b)}-p_{\mu_a}p_{\nu_b}q_{(\mu_a|\nu_b)}-p_{\nu_a}p_{\mu_b}q_{(\nu_a|\mu_b)}$. 
The average of the pairing interval can be given by [Eq. (C25) in Ref.~\cite{zhou2022experimental}]
\begin{equation}
\begin{aligned}\label{T_mean}
T_{\rm mean}=\frac{1-N_{T_{c}}q_{\rm tot}(1/q_{T_{c}}-1)}{Fq_{\rm tot}}.
\end{aligned} 
\end{equation}

The total number of set $\mathcal{S}_ {[k_a^{\rm tot}, k_b^{\rm tot}]}$ (except set $\mathcal{S}_{[2\nu_a,2\nu_b]}$) is  [Eq. (C26) in Ref.~\cite{zhou2022experimental}]
 \begin{equation}
\begin{aligned}
 n_{[k_a^{\rm tot}, k_b^{\rm tot}]}& =n_{\rm tot}\times \\
 \sum\limits_{k_a^e+k_a^l= k_a^{\rm tot}}&\sum\limits_{k_b^e+k_b^l=k_b^{\rm tot}}  
 \left(\frac{p_{k_a^e}p_{k_b^e}q_{(k_a^e|k_b^e)}}{q_{\rm tot}}\frac{p_{k_a^l}p_{k_b^l} q_{(k_a^l|k_b^l)}}{q_{\rm tot}}\right).
\end{aligned} 
\end{equation}
The total number of set $\mathcal{S}_{[2\nu_a, 2\nu_b]}$  is  [Eq. (C27) in Ref.~\cite{zhou2022experimental}]
 \begin{equation}
\begin{aligned}
 n_{[2\nu_a,2\nu_b]}=   \frac{n_{\rm tot}}{M\pi} \int_0^{2\pi}  \left(\frac{p_{\nu_a}p_{\nu_b}q_{(\nu_a|\nu_b)}^{\theta}}{q_{\rm tot}}\frac{p_{\nu_a}p_{\nu_b}q_{(\nu_a|\nu_b)}^{\theta}}{q_{\rm tot}}\right) d\theta.
\end{aligned} 
\end{equation}
The total number of errors in the $\boldsymbol{X}$ basis can be written as  [Eq. (C28) in Ref.~\cite{zhou2022experimental}]
 \begin{equation}
\begin{aligned}
m_{[2\nu_a,2\nu_b]}=&\frac{n_{\rm tot}}{M\pi}      p_{\nu_a}^2p_{\nu_b}^2\times \\
\int_0^{2\pi}\Bigg\{(1-&E_{\rm{\rm HOM}})\frac{   \left[q_{(\nu_a|\nu_b)}^{\theta,L}q_{(\nu_a|\nu_b)}^{\theta+\delta ,R}+q_{(\nu_a|\nu_b)}^{\theta,R}q_{(\nu_a|\nu_b)}^{\theta+\delta ,L}\right]}{q_{\rm tot}^2}\\
+ &E_{\rm{\rm HOM}}\frac{  \left[q_{(\nu_a|\nu_b)}^{\theta,L}q_{(\nu_a|\nu_b)}^{\theta+\delta,L}+q_{(\nu_a|\nu_b)}^{\theta,R}q_{(\nu_a|\nu_b)}^{\theta+\delta ,R}\right]}{{q_{\rm tot}^2}}\Bigg\} d\theta,
\end{aligned} 
\end{equation}
where $E_{\rm{HOM}}$ is the interference misalignment error rate,  and $\delta = T_{\rm mean}(2\pi\Delta\nu+ \omega_{\rm fib})$ is the phase misalignment resulting from the fiber phase drift rate $\omega_{\rm fib}$ and laser frequency difference $\Delta\nu$.  

\subsection{Simulation formulas for four-intensity MDI-QKD}\label{simulation:MDIQKD}
We denote the number and error number of detection event when
Alice sends intensity $k_a$ $(k_a\in\{\mu_a, \nu_a, \omega_a, o_a\})$, and Bob sends  $k_b$ $(k_b\in\{\mu_b, \nu_b, \omega_b, o_b\})$ in the $\boldsymbol{Z}(\boldsymbol{X})$ basis
as  $n_{k_ak_b}^{z(x)}$ and $m_{k_ak_b}^{z(x)}$, respectively.
The key rate of time-bin MDI-QKD is~\cite{curty2014finite,jiang2021higher} 
\begin{equation}
\begin{aligned}
R=& \frac{1}{N'} \left\{\underline{n}_{0}^z+\underline{n}_{11}^z\left[1-H_2\left(\overline{\phi}_{11}^z\right)\right]-\lambda_{\rm EC} \right.
\\ 
&\left. - \log_2\frac{2}{\varepsilon_{\rm cor}}-2\log_2\frac{2}{\varepsilon'\hat{\varepsilon} }-2\log_2\frac{1}{2\varepsilon_{\rm PA}}\right\},
\end{aligned}\label{mdi_keyrate_func}
\end{equation}
where $\lambda_{\rm EC}=n_{\mu_a\mu_b}^z fH_2\left(\frac{m_{\mu_a\mu_b}^z}{n_{\mu_a\mu_b}^z}\right)$.

Here we use the decoy-state analysis to consider the complete finite-key effects and apply the double-scanning method to  MDI-QKD~\cite{jiang2021higher}. The corresponding parameters in Eq.~\eqref{mdi_keyrate_func} can be given by
 \begin{equation}
\begin{aligned}
\underline{n}_{0}^{z*}= &  \max\left\{
\frac{e^{-\mu_a } p_{\mu_a }}{p_{ o_a }}\underline{n}_{o_a \mu_b}^{ z*}, \frac{e^{-\mu_b}p_{\mu_b}}{p_{ o_b } }\underline{n}_{\mu_a o_b }^{z *}\right\},\\
 \underline{n}_{11}^{z*}=&  \frac{\mu_a\mu_be^{-\mu_a-\mu_b} p_{\mu_a}p_{\mu_a}}{\nu_a\nu_b\omega_a\omega_b(\omega'-\nu')}\left(\underline{P}^{+*}-\overline{P}^{-*}+\underline{\hat{M}}^*-\overline{\hat{H}}^*\right), \\
\overline{t}_{11}^{x*}=&\frac{1}{\omega_a\omega_b\omega'e^{\nu_a+\nu_b}}\left(\hat{M}-\frac{\hat{H}}{2} \right),\\ 
\overline{t}_{11}^{z*}=&\frac{\mu_a\mu_be^{-\mu_a-\mu_b} p_{\mu_a}p_{\mu_a}}{\nu_a\nu_be^{-\nu_a-\nu_b} p_{\nu_a}p_{\nu_a}}\overline{t}_{11}^{x*},\\
\overline{\phi}_{11}^z=&\frac{\overline{t}_{11}^z}{\underline{n}_{11}^z},
\end{aligned}
\end{equation}
where
\begin{equation}
   \begin{cases}
   \omega'=\omega_a,\quad \nu'=\nu_a& \mbox{if} \quad \frac{\omega_a}{\omega_b}
   \le \frac{\nu_a}{\nu_b}, \\
   \omega'=\omega_b, \quad\nu'=\nu_b&\mbox{if}  \quad \frac{\omega_a}{\omega_b}
   > \frac{\nu_a}{\nu_b}.
   \end{cases}
  \end{equation}
and
\begin{equation}
\begin{aligned}
P^{+*}=& \omega_a\omega_b\omega' e^{\nu_a+\nu_b }\frac{(n_{\nu_a \nu_b}^{x}-m_{\nu_a \nu_b}^{x})^*}{p_{\nu_a}p_{\nu_b}}\\
&+\nu_a\nu_b\nu'e^{\omega_a}\frac{n_{\omega_a o_b}^{x*}}{p_{\omega_a}p_{o_b}}+\nu_a\nu_b\nu'e^{\omega_b }\frac{n_{o_a \omega_b  }^{x*}}{p_{ o_a}p_{\omega_b}},\\
P^{-*}=	&\nu_a\nu_b\nu'  e^{\omega_a+\omega_b}\frac{n_{ \omega_a \omega_b  }^{x*}}{p_{ \omega_a}p_{\omega_b}}  
 +\nu_a\nu_b\nu'  \frac{n_{o_a o_b }^{x*}}{p_{o_a}p_{o_b}}, \\
\hat{M}^*=&  \omega_a\omega_b\omega'e^{\nu_a+\nu_b}\frac{m_{\nu_a\nu_b}^{x*}}{p_{\nu_a}p_{\nu_b}},\\
\hat{H}^*=&\omega_a\omega_b\omega' \left(e^{\nu_b}\frac{n_{ o_a \nu_b  }^{x*}}{p_{o_a}p_{\nu_b}}+ 
	e^{\nu_a }\frac{n_{\nu_ao_b }^{x*}}{p_{\nu_a}p_{o_b}}
	-\frac{n_{o_ao_b }^{x*}}{p_{o_a}p_{o_b}}\right). 
\end{aligned}
\end{equation}
By scanning $(\hat{H},\hat{M})$, we can obtain the secret key rate 
\begin{alignat}{2}
\min\quad & R &{}& \\
\mbox{s.t.}\quad
&\underline{\hat{H}} \leq \hat{H}\leq \overline{\hat{H}}, \\ \nonumber
&\underline{\hat{M}} \leq \hat{M}\leq \overline{\hat{M}}.  \nonumber
\end{alignat}  
 Because of the dead time of the detector, only one of the four Bell states can be identified. In the simulation, we set
\begin{equation}
\begin{aligned}
n_{k_ak_b}^z=& N'p_{k_a}p_{k_b} p_d(1-p_d)^2e^{-\frac{k_a\eta_a+k_b\eta_b}{2}} 
\\ 
&\left\{I_0(\sqrt{k_a\eta_ak_b\eta_b}) -(1-p_d) e^{-\frac{k_a\eta_a+k_b\eta_b}{2}}\right. \\
+&\left.\left[1-(1-p_d)e^{-\frac{k_a\eta_a}{2}}\right]\left[1-(1-p_d)e^{-\frac{k_b\eta_b}{2}}\right]\right\},\\
m_{k_ak_b}^z=&N'p_{k_a}p_{k_b} p_d(1-p_d)^2e^{-\frac{k_a\eta_a+k_b\eta_b}{2}} 
\\ 
&\left\{ \left[I_0(\sqrt{k_a\eta_ak_b\eta_b}) -(1-p_d) e^{-\frac{k_a\eta_a+k_b\eta_b}{2}}\right]\right\}, 
\end{aligned}
\end{equation}
and
\begin{equation}
\begin{aligned}
n_{k_ak_b}^{x}=&N'p_{k_a}p_{k_b}y_{k_ak_b}^2\left[1+2y_{k_ak_b}^2\right.\\
&\left.-4y_{k_ak_b}I_0\left(\frac{\sqrt{ k_a\eta_ak_b\eta_b}}{2}\right)+I_0(\sqrt{ k_a\eta_ak_b\eta_b})\right],\\
m_{k_ak_b}^{x}=&N'p_{k_a}p_{k_b}y_{k_ak_b}^2\left\{1+y_{k_ak_b}^2\right.\\
&-2y_{k_ak_b}I_0\left(\frac{\sqrt{ k_a\eta_ak_b\eta_b}}{2}\right)\\
&\left.
 +E_{\rm HOM}\left[ I_0(\sqrt{ k_a\eta_ak_b\eta_b})-1\right.\right\},\\
\end{aligned}
\end{equation}
where we have $y_{k_ak_b}=(1-p_d)e^{-\frac{k_a\eta_a+k_b\eta_b}{4}}$ and $E_{\rm HOM}=0.04$. Note that in time-bin MDI-QKD, two pulses form one bit, i.e., $N'=N/2$.

\subsection{Simulation formulas for  four-intensity decoy-state QKD}\label{simulation:QKD}
The key rate of decoy-state QKD  is~\cite{yin2020tight,lim2014concise} 
\begin{equation}
\begin{aligned}
R=& \frac{1}{N} \left\{\underline{n}_{0}^z+\underline{n}_{1}^z\left[1-H_2\left(\overline{\phi}_{1}^z\right)\right]-\lambda_{\rm EC} \right.
\\ 
&\left. -6\log_2\frac{23}{\varepsilon_{\rm sec}}-2\log_2\frac{2}{\varepsilon_{\rm cor}}\right\},
\end{aligned}
\end{equation}
where $\lambda_{\rm EC} =(n_{\mu}^z+n_{\nu}^z)fH_2\left(\frac{m_\mu^z+m_\nu^z}{n_\mu^z+n_\nu^z}\right)$, and $n_k^{z(x)}$ and $m_k^{z(x)}$ are the number and error number
of  intensity pulse $k$ $(k\in\{\mu, \nu, \omega, o\})$
 measured in the $\boldsymbol{Z(X)}$ basis, respectively.
 
\begin{widetext}
First, we extend the decoy state analysis to finite-size cases. The number of vacuum events in the $\boldsymbol{Z}$ and $\boldsymbol{X}$ bases satisfy
\begin{equation}
\begin{aligned} 
\underline{n}_0^{z*}=&\frac{p_\mu e^{-\mu}+p_\nu e^{-\nu}}{p_o}\underline{n}_{o}^{z*}, 
\end{aligned}
\end{equation}
and
\begin{equation}
\begin{aligned} 
\underline{n}_0^{x*}=&\frac{p_\omega e^{-\omega}}{p_o}\underline{n}_{o}^{x*}, 
\end{aligned}
\end{equation}
respectively. 

The number of single-photon events in the $\boldsymbol{Z}$ and $\boldsymbol{X}$ bases are
\begin{equation}
\begin{aligned} 
\underline{n}_1^{z*}=& \frac{(p_\mu \mu e^{-\mu}+p_\nu \nu e^{-\nu})\mu}{\mu\nu-\nu^2}\left(\frac{e^{\nu}\underline{n} _\nu^{z*}}{p_\nu}-\frac{\nu^2}{\mu^2}\frac{e^{\mu}\overline{n}_\mu^{z*}}{p_\mu}-\frac{\mu^2-\nu^2}{\mu^2}\frac{\overline{n}_o^{z*}}{p_o}\right),
\end{aligned}
\end{equation}
and
\begin{equation}
\begin{aligned} 
\underline{n}_1^{x*}=&  \frac{p_\omega \omega e^{-\omega} \mu}{\mu\nu-\nu^2}\left(\frac{e^{\nu}\underline{n}_\nu^{x*}}{p_\nu}-\frac{\nu^2}{\mu^2}\frac{e^{\mu}\overline{n}_\mu^{x*}}{p_\mu}-\frac{\mu^2-\nu^2}{\mu^2}\frac{\overline{n}_o^{x*}}{p_o}\right),
\end{aligned}
\end{equation}
respectively. In addition, the number of bit errors $\overline{t}_1^x$ associated with the single-photon events in the $\boldsymbol{X}$ basis is also required. It is given by
\begin{equation}
\begin{aligned}\overline{t}_1^x=m_\omega^x-\underline{m}_0^x, 
\end{aligned}
\end{equation}
where $\underline{m}_0^{x*}=\frac{p_\omega e^{-\omega}}{p_o}\underline{m}_{o}^{x*}$. 
Second, the formula for the phase error rate of the single-photon
events in the $\boldsymbol{Z}$ basis can be written as
\begin{equation}
\begin{aligned}
\overline{\phi}_1^z=&\frac{\overline{m}_1^x}{\underline{n}_1^x}+\gamma\left(\underline{n}_{1}^z,\underline{n}_{1}^x,\frac{\overline{t}_1^x}{\underline{n}_1^x},\varepsilon_e\right).
\end{aligned}
\end{equation}
In the simulation, we set 

\begin{equation}
\begin{aligned} 
n_{k}^z=&\frac{Np_{k}}{2}\left[ 1-(1-p_d^z)^2e^{-kq_z\eta^z}\right]\left[ 1+(1-p_d^x)^2e^{-kq_x\eta^x}\right],\\
m_{k}^z=&\frac{Np_{k}}{2}\left[ 1+(1-p_d^x)^2e^{-kq_x\eta^x}\right]\left\{ (e_0-e_m^z) \left[ 1-(1-p_d^z)^2\right] e^{-kq_z\eta^z}+e_m^z \left[ 1-(1-p_d^z)^2e^{-kq_z\eta^z}\right] \right\},
\end{aligned}
\end{equation}
and 
\begin{equation}
\begin{aligned}
n_{k}^x=&\frac{Np_{k}}{2}\left[ 1-(1-p_d^x)^2e^{-kq_x\eta^x}\right]\left[ 1+(1-p_d^z)^2e^{-kq_z\eta^z}\right],\\
m_{k}^x=&\frac{Np_{k}}{2}\left[ 1+(1-p_d^z)^2e^{-kq_z\eta^z}\right]\left\{(e_0-e_m^x)  \left[ 1-(1-p_d^x)^2\right] e^{-kq_x\eta^x} +e_m^x \left[ 1-(1-p_d^x)^2e^{-kq_x\eta^x}\right] \right\},
\end{aligned}
\end{equation}
where $e_0=1/2$  is the error rate of the background
noise, $e_m^z=e_m^x=e_m$, $p_d^z=p_d^x=p_d$, and $\eta^z=\eta^x=\eta_d 10^{-\frac{\alpha l+\eta_{\rm int}}{10}}$. The code of decoy-state QKD and decoy-state MDI-QKD has been uploaded to the open-source code website~\cite{xiecode}.
\end{widetext}
\section{Statistical fluctuation analysis}\label{statistical}
In this Appendix, we introduce the statistical fluctuation analysis method~\cite{yin2020tight} used in the simulation.

{\it{~Chernoff bound.}}
For a given expected value $x^*$ and failure probability $\varepsilon$, we can use the Chernoff bound to estimate the upper and lower bounds of the observed value
\begin{equation}
\begin{aligned}\label{chernoff1}
\overline{x}&=\varphi^U(x^{*})=x^{*}+\frac{\beta}{2}+\sqrt{2\beta x^{*}+\frac{\beta^{2}}{4}},
\end{aligned}
\end{equation}
and
\begin{equation}
\begin{aligned}\label{chernoff2}
	\underline{x}&=\varphi^L(x^{*})=x^{*}-\sqrt{2\beta x^{*}},
\end{aligned}
\end{equation}
where $\beta=\ln{\epsilon^{-1}}$.

{\it{~Variant of Chernoff bound.}}
The variant of the Chernoff bound can help us estimate the expected value from their observed values. One can apply the following equations to  obtain the upper and lower bounds of $x^{*}$ 
\begin{equation}
\begin{aligned}\label{varchernoff1}
	\overline{x}^{*}&=x+\beta+\sqrt{2\beta x+\beta^{2}},\\
\end{aligned}
\end{equation}
and
\begin{equation}
\begin{aligned}\label{varchernoff2}
	\underline{x}^{*}&=\max\left\{x-\frac{\beta}{2}-\sqrt{2\beta x+\frac{\beta^{2}}{4}},~0\right\}.\\
\end{aligned}
\end{equation}


 %

\end{document}